\documentclass[numbers]{elsarticle}
\usepackage{latexsym}
\usepackage{float}
\usepackage[latin2]{inputenc}
\usepackage{graphicx}
\usepackage{ae}
\usepackage{aecompl}
\usepackage{setspace}
\usepackage{booktabs}
\usepackage[hidelinks]{hyperref}
\usepackage{lineno}

\makeatletter
\def\ps@pprintTitle{%
	\let\@oddhead\@empty
	\let\@evenhead\@empty
	\def\@oddfoot{\centerline{\thepage}}%
	\let\@evenfoot\@oddfoot}
\makeatother

\title{Opening Amyloid-Windows to the Secondary Structure of Proteins: The Amyloidogenecity Increases Tenfold Inside Beta-Sheets}
		
		\author[p]{Kristóf Takács}
		\ead{takacs@pitgroup.org}
		\author[p]{Bálint Varga}
		\ead{balorkany@pitgroup.org}
       	\author[s]{Viktor Farkas}
		\ead{farkas.viktor@ttk.elte.hu}
		\author[s,t]{András Perczel}
		\ead{perczel.andras@ttk.elte.hu}
		\author[p,u]{Vince Grolmusz\corref{cor1}}
		\ead{grolmusz@pitgroup.org}
		\cortext[cor1]{Corresponding author}
		\address[p]{PIT Bioinformatics Group, Eötvös University, H-1117 Budapest, Hungary}
		\address[u]{Uratim Ltd., H-1118 Budapest, Hungary}
		\address[s]{ELKH-ELTE Protein Modeling Research Group, H-1117 Budapest, Hungary}
		\address[t]{Laboratory of Structural Chemistry and Biology, Eötvös University, H-1117, Budapest, Hungary}

\begin{document} 	

\begin{abstract}
Methods from artificial intelligence (AI), in general, and machine learning, in particular, have kept conquering new territories in numerous areas of science. Most of the applications of these techniques are restricted to the classification of large data sets, but new scientific knowledge can seldom be inferred from these tools. Here we show that an AI-based amyloidogenecity predictor can strongly differentiate the border- and the internal hexamers of $\beta$-pleated sheets when screening all the Protein Data Bank-deposited homology-filtered protein structures. Our main result shows that more than 30\% of internal hexamers of $\beta$ sheets are predicted to be amyloidogenic, while just outside the border regions, only 3\% are predicted as such. This result may elucidate a general  protection mechanism of proteins against turning into amyloids: if the borders of $\beta$-sheets were amyloidogenic, then the whole $\beta$ sheet could turn more easily into an insoluble amyloid-structure, characterized by periodically repeated parallel $\beta$-sheets. We also present that no analogous phenomenon exists on the borders of $\alpha$-helices or randomly chosen subsequences of the studied protein structures.
\end{abstract}

\date{}
	
\maketitle

\section*{Introduction} 

The amyloid status of proteins are studied for a long time because of its relevance in human diseases, including neurodegenerative ailments \cite{Nelson2005,Eisenberg2012,Horvath2019,Taricska2020}, amyloidoses \cite{Amyloidosis} and other diseases, listed in Table 1 of \cite{Eisenberg2012}. Recently, the amyloids gained increased attention as possible antiviral agents \cite{Michiels2020}, or in one of the first successes in the human {\em in vivo} CRISPR-Cas9-based gene editing therapy of the transthyretin amyloidosis \cite{Gillmore2021}. 

The insoluble amyloid state differs from the unstructured protein aggregates; it has a well-defined structure characterized by repetitive parallel $\beta$-sheets \cite{Takacs2019,Takacs2020}. 

Numerous globular proteins may turn into insoluble amyloid structures in certain pH and temperature combinations. It is observed that the transition to the amyloid state is similar to the contagious prion-formation: it is related to a core-forming structure \cite{Eisenberg2012} for starting the amyloid-transition, and the subsequent propagation of misfolded, insoluble, parallel $\beta$-sheets \cite{Ban2006,Knowles2011}. The amyloid propagation is related to the spatially accessible, neighboring $\beta$-pleated sheet secondary structural elements of proteins \cite{Chiti2017,Horvath2019,Taricska2020}.

Naturally occurring, globular, soluble proteins need to possess certain structural properties which prevent them from turning into insoluble amyloid aggregates. Experimental studies, together with {\it in silico} methods able to predict aggregation-prone regions (APRs) in protein sequences. In the literature, certain residues acting as ``gatekeepers'' are described as preventing those conformational changes to amyloid structures: aspartic acid and glycine in bacterial curlin subunits CsgA and CsgB \cite{Wang2010a}, lysine in transthyretin \cite{SantAnna2014} or in peptides Trpzip1 and Trpzip2 \cite{Markiewicz2014}. 

Our hypothesis is that the borders of $\beta$-sheets in globular, soluble proteins in general (and not only in the specific examples listed above) need to be protected against turning to amyloid structures; otherwise the whole polypeptide chain bordering the $\beta$-sheet subsequences would be transformed to amyloids. We examine this hypothesis in the present work by screening the starting and ending subsequences, or in other terminology, the prefixes, and suffixes of the $\beta$-sheets of the homology-filtered Protein Data Bank (PDB) entries.

\subsection*{Previous work}

In \cite{Takacs2019}, a clean geometrical formulation was given for the characterization of the amyloid-like structures in length, distance, and parallelism of their $\beta$-sheets, and by the search of these geometric properties, an automatically updated list of the PDB-entries was published and maintained at the address \url{https://pitgroup.org/amyloid/}. 

In \cite{Takacs2020}, the amyloid-precursor molecules of \url{https://pitgroup.org/amyloid/} were classified by the prefixes of their parallel $\beta$-sheet subsequences, and these prefixes were also searched for in the whole Protein Data Bank \cite{PDB-base}. The remarkable findings include the major histocompatibility complexes MHC-1 and MHC-2, the p53 tumor suppressor protein, and an anti-coagulant peptide molecule. 

With a prediction tool, based on artificial intelligence, in \cite{Keresztes2020a}, we partitioned all the possible hexapeptides into two classes: ``amyloidogenic'' and ``non-amyloidogenic'' with more than 84\% correctness. The prediction tool, which applies a linear Support Vector Machine (SVM) \cite{Cortes1995}, was trained by the Waltz dataset of 514 amyloidogenic and 901 non-amyloidogenic hexapeptides \cite{Beerten2015, Louros2020}. The corresponding web-based tool is available as the Budapest Amyloid Predictor at \url{https://pitgroup.org/bap}. 

As a surprising application of the Budapest Amyloid Predictor, numerous amyloid- and non-amyloid hexapeptide patterns were identified in \cite{Keresztes2022}. For example, it was shown that for all independent amino acid substitutions for the positions marked by ``x'', the CxFLWx or FxFLFx patterns describe amyloidogenic hexamers, while the patterns PxDxxx or xxKxEx describe non-amyloidogenic hexamers. Note that each pattern with two x's describes $20^2=400$, while those with four x's describe $20^4=160,000$ different hexapeptides succinctly if for the positions x, we can substitute the 20 amino acid residues. 

In the present work, we study the prefixes and suffixes of the $\beta$-sheet subsequences of the PDB-deposited structures and apply the Budapest Amyloid Predictor to the hexamers of the border regions on these $\beta$-sheets. As we show below, the number of the amyloidogenic hexapeptides of the border regions of the $\beta$-sheets is one-tenth of the same number in the inner parts of those sequences by screening the homology-filtered Protein Data Bank entries. We believe that this finding is a very strong evidence for our hypothesis.

We remark that applying the artificial intelligence-based Budapest Amyloid Predictor for the 64 million ($=20^6$) possible hexapeptides is a crucial step here: the largest experimental dataset that labels the hexapeptides by their amyloidogenic propensity, the Waltz dataset, contains only 1415 hexapeptides \cite{Beerten2015, Louros2020}, while we analyzed the prefixes and suffixes of more than 110,000 $\beta$-sheets from the homology-filtered PDB. Gaining experimental data for that many hexapeptides is intractable today. 

\section*{Methods}

For the statistical analysis involving the entries of the Protein Data Bank (PDB) \cite{PDB-base}, usually some ``non-redundant'' homology-filtered PDB-subsets are applied because otherwise, the multiple depositions of more important or more researched protein structures (e.g., several thousand copies of the HIV-1 protease) would impact the results. In this study, we have made use of the representative polypeptide chains of the 30\% homology-filtered set, which was available at the URL \url{https://www.rcsb.org/pdb/rest/representatives?cluster=30} (downloaded on May 21, 2020). 

{\it Technical note for reproducibility of the present work: At the time of writing this section, we were informed that the administrators of the RCSB PDB have removed the link target indicated above, but one can still generate the similar non-redundant set with an advanced search feature of the RCSB PDB, namely \url{https://search.rcsb.org/#group-by-context}, by choosing 30\% sequence identity.}

The structure of almost all representative members of this non-redundant set was identified in a soluble state, so almost all proteins in the set are water-soluble (see Figure S3 in the supporting material).

From the representative sequences of the non-redundant set, we have identified the maximal contiguous subsequences corresponding to $\beta$-sheets by applying the ``SHEET'' records in the PDB file. 
In other words, all the non-expandable contiguous $\beta$-sheet subsequences were identified. Those subsequences whose lengths are less than 6 were discarded. This way, we have identified as many as 117,467 subsequences. 

In the next step, we listed the six-residue long sliding windows through the borders of these 117,467 $\beta$-sheet subsequences, as it is visualized in Figure 1.

\begin{figure}[H]
	\begin{center}
		\includegraphics[width=12cm]{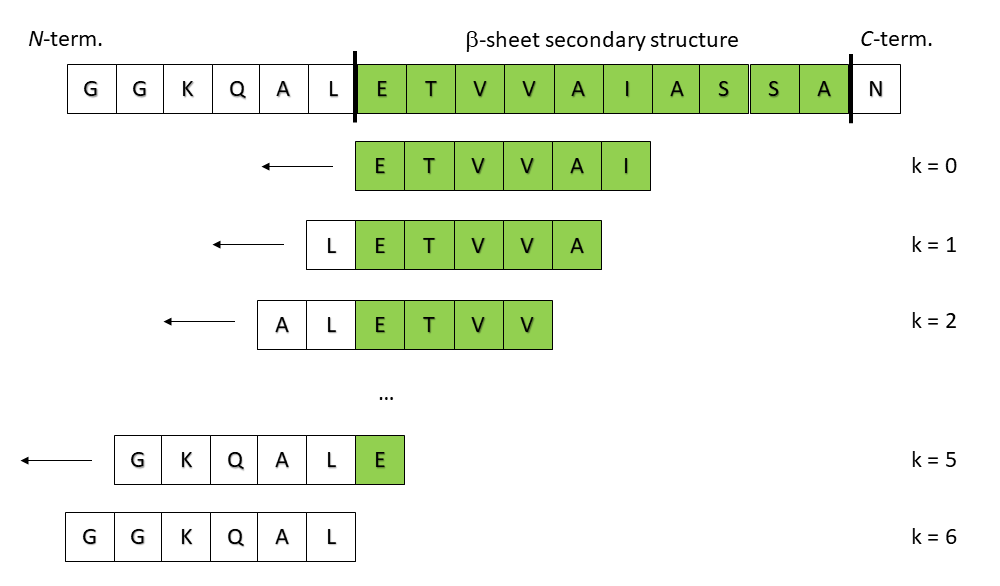}
\caption{A visualization of the sliding prefix windows of $\beta$-sheet units, mapping their border regions.  A section of the polypeptide sequence of a protein is drawn at the top of the figure, listed in the standard N-terminal through C-terminal order. The green shaded subsequence represents the $\beta$-sheet region, identified from the ``SHEET'' record of the PDB file. The rows, annotated by $k=0,1,\ldots,6$, correspond to the sliding windows of length 6 residues, when the window slides to the left, and in row, labeled by $k=i$, exactly $i$ residues are outside the green-shaded $\beta$-sheet region, for $i=0,1,2,\ldots,6$. In the case of sliding suffix windows (Figure S1 in the supporting material), the windows move to the right, and, similarly to the prefix case, $k=0$ corresponds to the window entirely inside the right end of the green subsequence, and $k=6$ to the configuration where the length-6 window is just outside to the right of the green sequence in its entirety.}
	\end{center}
\end{figure}

The six residue-long windows map the border regions of the $\beta$-sheets, identified from the ``SHEET'' record of the PDB file. For each contiguous $\beta$-sheet section, we consider the sliding windows of its left border (Figure 1), called prefixes, and the seven sliding windows at the right border, called suffixes. That is, for each of the 117,467 unique subsequences, we have listed the seven prefix and the seven suffix sequences (for $k=0,1,2,3,4,5,6$), and applied the Budapest Amyloid Predictor \cite{Keresztes2020a} at \url{https://pitgroup.org/bap} for assigning one of the ``amyloidogenic'' or ``non-amyloidogenic'' labels for each of the seven length-6 windows at both ends of the 117,467 subsequences. 

Next, we have drawn a diagram, Figure 2, which contains at the $x$ axis the $k$ value (i.e., the length of the window over (or outside) the $\beta$-sheet sequence (colored by green in Figure 1). On the $y$ axis, we give the percentage of the predicted amyloidogenic length-6 windows, computed by the Budapest Amyloid Predictor. In other words, the $y$ value gives the fraction of the hexapeptides with exactly $k$ residues outside the $\beta$-sheet region, predicted to be amyloidogenic.

\begin{figure}[H]
	\begin{center}
		\includegraphics[width=12cm]{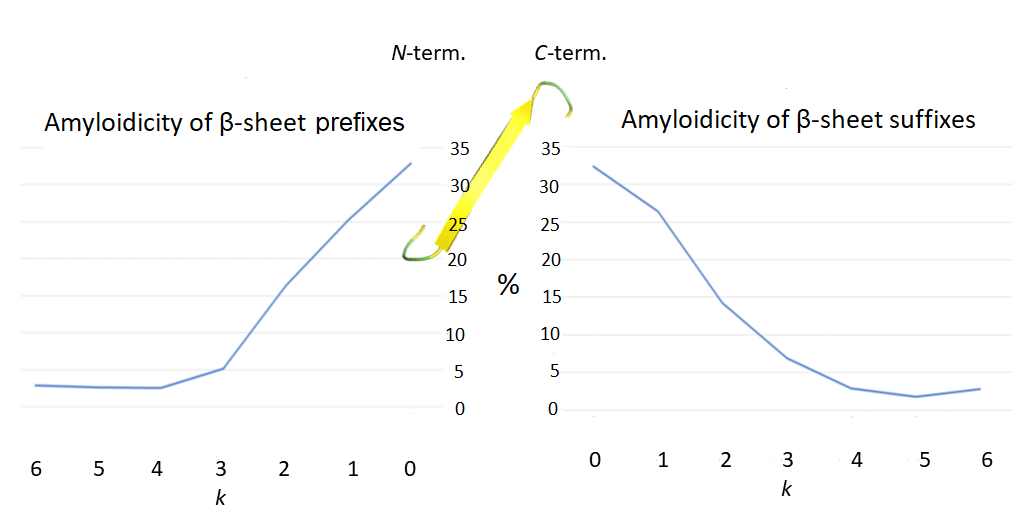}
		\caption{The percentage of the length-6 prefix windows (left panel) and suffix windows (right panel) with $k$ positions outside the $\beta$-sheet region, predicted to be amyloidogenic by the Budapest Amyloid Predictor. The $x$ axis shows the $k$ value (i.e., the length of the window over (or outside) the $\beta$-sheet sequence (colored by green in Figure 1 and S1, resp.), and on the $y$ axis, the percentage of the amyloidogenic length-6 windows are given. In the left panel, the minimum value is 2.54\% at $k=4$, at $k=6$ the percentage is 2.89, the maximum is taken at $k=0$ with 32.9\%. In the right panel, the minimum value is 1.79\% at $k=5$, at $k=6$ the percentage is 2.81, the maximum is taken at $k=0$ with 32.4\%. Both graphs show that by crossing the borders of the $\beta$-sheet-regions, the amyloidocity percentage of the hexapeptide windows drops to less than 1/10 of the starting value. Or, in other words, more than 10 times more hexapeptide windows are amyloidogenic just inside the $\beta$-sheet region than just outside of it. }
	\end{center}
\end{figure}

\section*{Discussion and results} 

The rapid development of artificial intelligence tools may open up new insights into the structural properties of biological macromolecules. For example, the largest available hexapeptide database, the Waltz dataset \cite{Beerten2015, Louros2020} today contains 1415 hexapeptides, labeled as ``amyloidogenic'' or ``non-amyloidogenic'', based on different types of experimental data. The total number of all the hexapeptides from the biologically most relevant 20 amino acids is $20^6=64 000 000$. Therefore, it is improbable that experimental evidence could be gained for the amyloid-forming properties of all of these hexapeptides. 

By artificial intelligence tools, however, we were able to build the Budapest Amyloid Predictor, which predicts the amyloidocity of any hexapeptide from these 64 million with an 84\% success rate \cite{Keresztes2020a}, \url{https://pitgroup.org/bap}.

The BAP predictor makes it possible to assign amyloidogenecity labels to any hexapeptides, not only the experimentally identified ones. Consequently, we may screen all the given secondary structures deposited in the PDB and compute the prediction for any hexapeptides in them. These actions were not possible without highly developed artificial intelligence tools.

Our results are summarized in Figure 2.

Figure 2 shows a monotonic tenfold increase of the amyloidogenecity of the hexapeptides, described by the sliding windows, for positions entirely in the $\beta$-sheet sequences ($k=0$), relative to the other positions, when the largest part of the windows are outside of the $\beta$-sheet region $k=4,5,6$. This observation validates our hypothesis, described in the Introduction, namely, the prefixes and the suffixes of the $\beta$-sheet subsequences need to be protected against turning into amyloid structures Otherwise, the whole polypeptide chain bordering the $\beta$-sheet subsequences would be transformed to amyloids. 

\subsection*{The $\alpha$-helix control}

For a simple control of our hypothesis, we also prepared the sliding prefix- and suffix windows for the $\alpha$-helices, found by screening all the homology-filtered PDB entries. As the supporting Figure S2 shows in the Supporting Material, no similar observation can be done in the case of the $\alpha$-helices.  

\subsection*{The random control} 

For another simple control, let us consider random (contiguous) subsequences of the PDB-deposited proteins instead of subsequences of $\beta$-sheets or $\alpha$-helices and evaluate their prefixes or suffixes for amyloidogenecity. The following reasoning shows that in this case, the amylodocity percentage of prefixes or suffixes of different $k$ values are almost the same (the ``almost'' adjective is explained below). We consider prefixes here; the proof for suffixes is analogous.

We refer to Figure S4 in the Supporting Material for visualizing the reasoning.

First, assume that we consider only fixed-length subsequences (of length $\ell\geq 6$). Let us consider a hexapeptide, which is not too close to the right end of the protein sequence, more exactly, whose first position is at least 6+$\ell$ residues from the right end (denoted by yellow in the first row of Fig. S4). 

It is easy to see that the {\it very} same (yellow) hexapeptide may be a prefix-window with $k=0$ for the green, with $k=1$ for the blue, with $k=2$ for the red, or with $k=3$ for the brown subsequence of length $\ell$. That is, if a hexapeptide is not too close to the right end of a protein chain, then it appears with $k=1,2,3,4,5,6$ for different length $\ell$ sequences. Let us remark that the amyloidocity of a hexapeptide depends only on its sequence, so there would not be any difference of prefix-amyloidocities for different $k$ values, unlike in the case of $\beta$-sheets in Fig. 2. This reasoning works also for random subsequences of different lengths, and also for suffixes. Small variability of the appearance percentage of a given hexapeptide may occur only if it is close to the right end of the protein-sequence and, consequently, cannot appear with larger $k$ or $\ell$ values.

\section*{Conclusions}
Here we screened all the prefix- and suffix sequences of length-6 of the $\beta$-sheets in the homology-filtered Protein Data Bank and have found that by crossing the boundaries of the $\beta$-sheet sequences the amyloidogenic ratio of the hexapeptides, characterized by the length-6 sliding windows, moving from outside to inside of the $\beta$-sheet sequence, increases more than tenfolds. 

No similar phenomenon can be observed in the case of $\alpha$-helices or random subsequences. We strongly believe that this fact shows the validity of our hypothesis, namely, the prefixes and the suffixes of the $\beta$-sheet subsequences need to be strongly protected against turning into amyloid structures, otherwise the whole polypeptide chain, bordering the $\beta$-sheet subsequences would possibly be transformed into amyloids. 
	
\section*{Data availability} The Budapest Amyloid Predictor webserver is available freely at \url{https://pitgroup.org/bap}.

\section*{Funding}
KT, BV, and VG were partially supported by the  NKFI-127909 grant of the National Research, Development and Innovation Office of Hungary. KT, BV, AP, VF, and VG were partially supported by the ELTE Thematic Excellence Programme (Szint+) subsidized by the Hungarian Ministry for Innovation and Technology.

\section*{Author Contribution} AP, VF, and VG initiated the study and evaluated results; KT collected \& identified the sequences to be analyzed, generated the sliding windows and analyzed the results, created the figures, and provided the control studies; BV assigned the amyloidogenecity labels to hexapeptides, VG has overseen the work and wrote the first version of the paper. AP, VF, and VG secured funding \& improved the manuscript.

\section*{Conflicting interest} The authors declare no conflicting interests.

%\section*{References}

%\bibliography{v:/vince/CIKKEK/medl}
%\bibliographystyle{unsrtnat}

\section*{Appendix: Supplementary material}

\newcommand{\beginsupplement}{
    \setcounter{section}{0}
    \renewcommand{\thesection}{S\arabic{section}}
    \setcounter{equation}{0}
    \renewcommand{\theequation}{S\arabic{equation}}
    \setcounter{table}{0}
    \renewcommand{\thetable}{S\arabic{table}}
    \setcounter{figure}{0}
    \renewcommand{\thefigure}{S\arabic{figure}}
    \newcounter{SIfig}
    \renewcommand{\theSIfig}{S\arabic{SIfig}}}

\beginsupplement

We present here Supplementary Figures for our work.

Figure S1 is the suffix-version of Figure 1 in the main text: it visualizes the sliding suffix windows of length 6.
\medskip

Figure S2 is the $\alpha$-helix version of Figure 2 in the main text; it yields a control for our main result.
\medskip

Figure S3 visualizes the percentage of the water-soluble proteins in the non-redundant PDB set we worked with. 
\medskip

Figure S4 explains the random control of our main result.

\begin{figure}[H]
	\begin{center}
		\includegraphics[width=11cm]{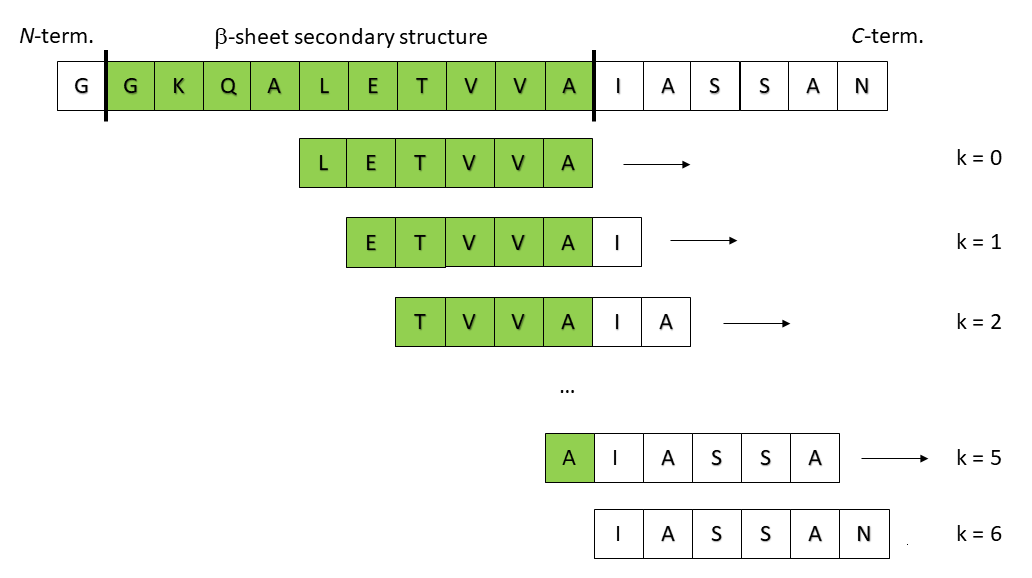}

		\caption{A visualization of the sliding suffix windows of $\beta$-chains, mapping their border regions.  A section of the polypeptide sequence of a protein is drawn at the top of the figure, listed in the standard N-terminal through C-terminal order. The green shaded subsequence represents the $\beta$-sheet region, identified from the ``SHEET'' record of the PDB file. The rows, annotated by $k=0,1,\ldots,6$, correspond to the sliding windows of length 6 residues, when the window slides to the left, and in row, labeled by $k=i$, exactly $i$ residues are outside the green-shaded $\beta$-sheet region, for $i=0,1,2,\ldots,6$.}
	\end{center}
\end{figure}

\begin{figure}[H]
	\begin{center}
		\includegraphics[width=11cm]{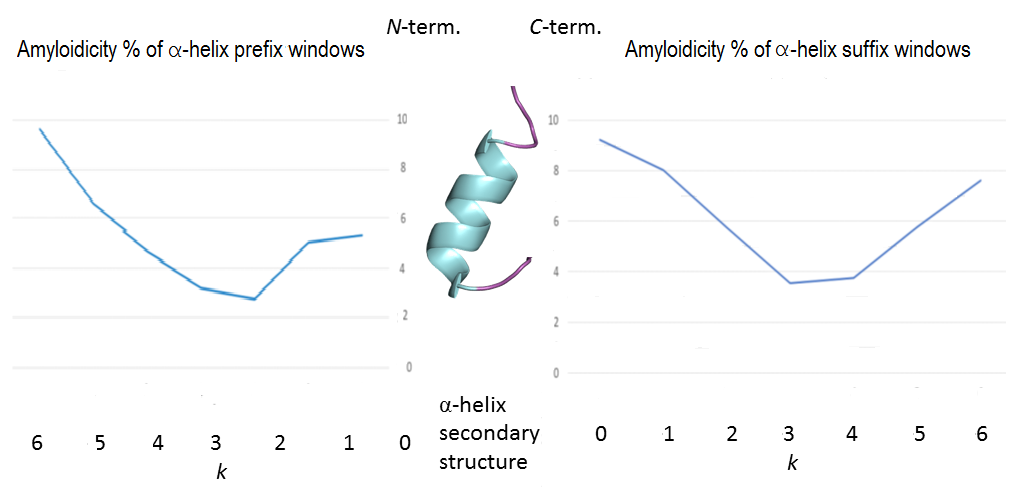}
		\caption{The percentage of the amyloidogenic length-6 prefix windows (left panel) and suffix windows (right panel) with $k$ positions outside the $\alpha$-helix region, predicted by the Budapest Amyloid Predictor. The $x$ axis shows the $k$ value (i.e., the length of the window over (or outside) the $\alpha$-helix sequence, and on the $y$ axis, the percentage of the amyloidogenic length-6 windows are given. The difference between the small and large $k$ values is much less than in the $\beta$-sheet case. }
	\end{center}
\end{figure}

\begin{figure}[H]
	\begin{center}
		\includegraphics[width=11cm]{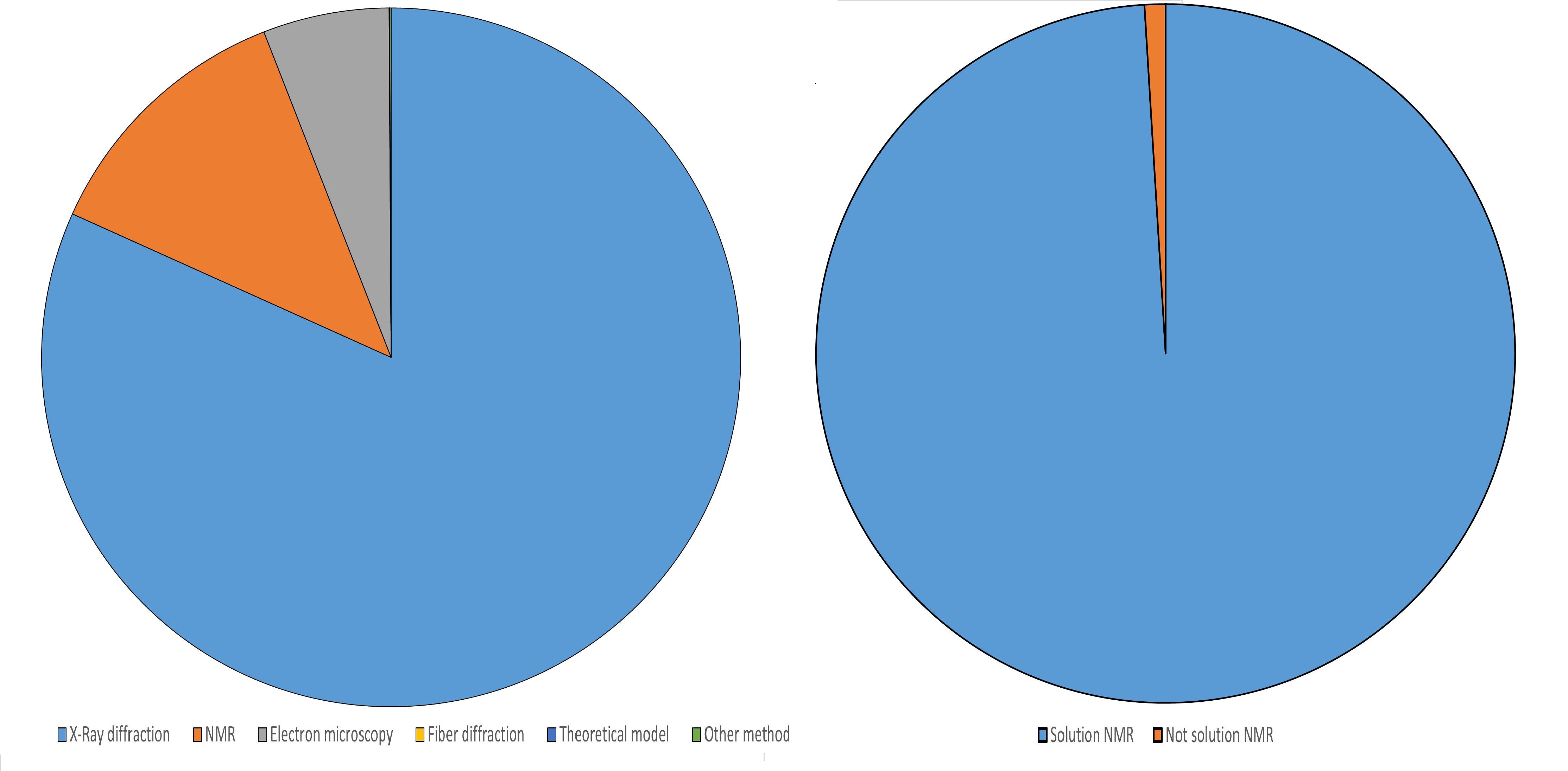}
		\caption{The distribution of the protein-structure identifications methods in the non-redundant set we have analyzed. The left panel contains the overall distribution of the methods and the right panel refines the NMR methods, differentiating soluble and solid-state methods. Clearly, almost all proteins were water soluble in the set.}
	\end{center}
\end{figure}

\begin{figure}[H]
	\begin{center}
		\includegraphics[width=11cm]{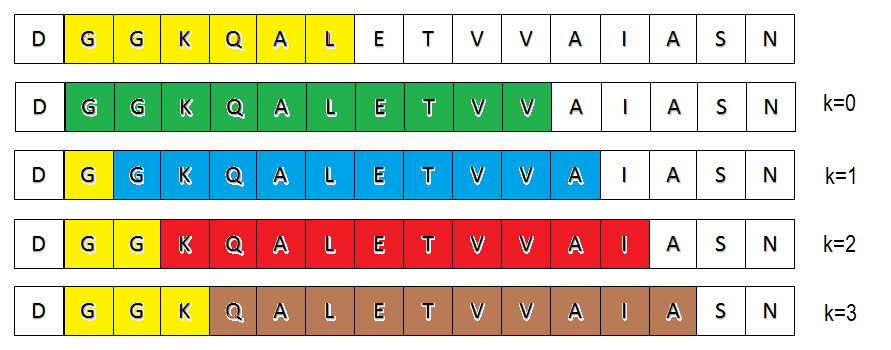}
		\caption{Random subsequence control for our main result. Consider the yellow hexapeptide in the first row of the figure. It is easy to see that the same (yellow) hexapeptide may be a prefix-window with $k=0$ for the green, with $k=1$ for the blue, with $k=2$ for the red, or with $k=3$ for the brown subsequence of length $\ell$. That is, if a hexapeptide is not too close to the right end of a protein chain, then it appears with $k=1,2,3,4,5,6$ for different length $\ell$ sequences. Let us remark that the amyloidocity of a hexapeptide depends only on its sequence, so there would not be any difference of prefix-amyloidocities for different $k$ values, unlike in the case of $\beta$-sheets in Fig. 2.}
	\end{center}
\end{figure}

\end{document}